\documentclass{mem}
\usepackage{natbib}\usepackage{txfonts}\usepackage{balance}
\usepackage{graphicx}
\usepackage[a4paper]{hyperref}
\idline{75}{282}
\begin{document}
\def\teff{$T\rm_{eff }$}

\title{Are There Radical Cyanogen Abundance Differences Between Galactic
Globular Cluster RGB and AGB Stars?}

\subtitle{Possibly a Vital Clue to the Globular Cluster Abundance Anomaly Problem}

\author{
S.W. \,Campbell\inst{1}, 
J.C. \,Lattanzio\inst{1}\and
L.M. \,Elliott\inst{1}
          }

  \offprints{S.W. Campbell}

\institute{Centre for Stellar and Planetary Astrophysics, Monash University, Melbourne, Australia.
           \email{simon.campbell@sci.monash.edu.au}}

\authorrunning{Campbell \& Lattanzio}

\titlerunning{Abundance Differences in GCs}


\abstract{

On reading an old paper about galactic globular cluster abundance observations
(of NGC 6752) we came across an intriguing result. \citet{NCF81} 
found that there was a distinct lack of cyanogen-strong
(CN-strong) stars in their sample of AGB stars, as compared to their
sample of RGB stars (which had roughly equal numbers of CN-normal and
CN-strong stars). Further reading revealed that similar features have
been discovered in the AGB populations of other
clusters. Recently, \citet{SIK00} followed up on this possibility
(and considered other proton-capture products) by compiling the existing
data at the time and came to a similar conclusion for two more clusters.
Unfortunately all of these studies suffer from low AGB star counts
so the conclusions are not necessarily robust --- larger, statistically 
significant, sample sizes are needed. 

In this conference paper, presented at the Eighth Torino
Workshop on Nucleosynthesis in AGB Stars (Universidad de Granada, Spain, 2006), we outline the results
of a literature search for relevant CN observations and describe our
observing proposal to test the suggestion that there are substantial
abundance differences between the AGB and RGB in galactic globular
clusters. The literature search revealed that the AGB star counts
for all studies (which are not, in general, studies about AGB stars in particular)
are low, usually being $\leq 10$. The search also revealed
that the picture may not be consistent between clusters. Although most clusters 
appear to have CN-weak AGBs, at least two seem to have CN-strong AGBs (M5 \& 47 Tuc).
To further complicate the picture, clusters often appear to have a combination
of both CN-strong and CN-weak stars on their AGBs -- although one population 
tends to dominate. Again, all these assertions 
are however based on small sample sizes. We aim to increase the sample sizes
by \textit{an order of magnitude} using existing high quality photometry in which
the AGB and RGB can be reliably separated.
For the observations we will use a wide-field, low- to mid-resolution multi-object
spectroscope to obtain data not only on the AGB but also on the horizontal 
branches and first giant branches of a sample of clusters. With the new
information we hope to ascertain whether significant abundance differences really exist.

          } 

\maketitle{}


\section{Introduction}

We are attempting to perform a conclusive test of the suggestion put
forward by \citet{NCF81}, which has been touched upon by many authors
since and recently explored by \citet{SIK00},
that there are differences in cyanogen abundance distributions between
the first and second giant branches in galactic globular clusters. 

Although galactic globular clusters (GCs) are chemically homogeneous
with respect to Fe and most other heavy elements (see eg. \citet{KSL92}),
it has long been known that they show inhomogeneities in many lighter
elements (eg. C, N, O, Mg, Al). These inhomogeneities are considered
anomalous because they are not seen in halo field stars of similar
metallicity (see eg. \citet{GSC00}). 

One of the first inhomogeneities discovered was that of the molecule
Cyanogen (CN, often used as a proxy for nitrogen). A picture of `CN-bimodality'
emerged in the early 1980s whereby there appears to be two distinct
chemical populations of stars in most, if not all, GCs. One population is known
as `CN-strong', the other `CN-weak' (the CN-weak population
might be more informatively called 'CN-normal' -- as these stars show
CN abundances similar to the Halo field stars). Originally, observations
of CN were mainly made in stars on the giant branches
but more recently there have been observations on the
main sequence (MS) and sub-giant branch (SGB) of some clusters (eg.
\citet{CCB98}). These observations show that there is little
difference in the bimodal CN pattern on the MS and SGB as compared
with the giants --- indicating a primordial origin for the differing
populations. Figure 6 in \citet{CCB98} exemplifies this situation.

Due to the paucity of asymptotic giant branch (AGB) stars in GCs (a
result of their short lifetimes) there have been very few systematic
observational studies of the CN anomaly on the AGB in globular clusters 
(\citet{M78} is one that the Authors are aware of).
What little that has been done has been an aside in more general papers
(eg. \citet{NCF81}, \citet{BSH93}, \citet{ISK99}). However
these studies have hinted at a tantalising characteristic: most (observed)
GCs show a lack of CN-strong stars on the AGB. If this is true then
it is in stark contrast to the red giant branch (RGB) and earlier phases 
of evolution, where the ratio of CN-Strong to CN-Weak stars is roughly unity in many clusters. 

This \textit{possible} discrepancy was noted
by \citet{NCF81} in their paper about abundances in
giant stars in NGC 6752. They state that ``The behaviour of the CN
bands in the AGB stars is... quite difficult to understand... not
one of the stars studied here has enhanced CN... yet on the [first]
giant branch there are more CN strong stars than CN weak ones.''
(also see Figure 3 in that paper). More recently \citet{SIK00}
presented a conference paper on this exact topic. Compiling the contemporaneous
preexisting data in the literature they discussed the relative amounts
of CN in AGB and RGB stars in the GCs NGC 6752 (data from \citet{NCF81}, 
M13 (data from \citet{S81}) and M4 (data from \citet{NCF81} and 
\citet{SS91}). They also discuss
Na abundance variations in M13 (data from \citet{PSK96a} and \citet{PSK96b}
). Their conclusion for the CN variations was that the clusters
in question all showed significantly less CN on the AGB as compared
to the RGB. However the data compiled only contained about 10 AGB
stars per cluster. In their closing remarks they suggest observations
with larger sample sizes are needed --- which may be done using wide-field
multi-object spectroscopes. This is exactly the conclusion the present
authors also came to, inspiring this seminar/conference paper at the Eighth Torino
Workshop on Nucleosynthesis AGB Stars held at the Universidad de Granada, Spain, in 2006.


\section{Literature Search Results and the Observing Proposal}

We conducted a literature search (which may not be complete)
to ascertain what work had already been done in terms on CN on the AGB
in galactic globular clusters. The results are displayed in Table 
\ref{table1}. The main result from this search was that the available number of
AGB star observations are not statistically significant enough to come
to any real conclusion about the nature of the CN abundance distributions.
This has mainly been due to technological constraints. However, the data that
does exist shows that there appears to be a strong trend towards CN-weak 
asymptotic giant branches.
The picture is not so simple though, as two clusters in Table \ref{table1} actually have
CN-strong AGBs. In addition to this, most clusters have some stars of the 
opposite class on their AGBs -- the classifications given in Table \ref{table1} 
(usually) refer to strong majorities in each cluster, rather than totally 
homogeneous populations.

\begin{table*}
\caption{Results of the literature search for CN abundances in GC AGB stars. Note that `weak' 
or `strong' means that there is a very significant majority of that class of star in each case.}
\label{table1}
\begin{center}
\begin{tabular}{|c|c|c|l|}
\hline
  \bf{Cluster}&
  \bf{No. AGB Stars}&
  \bf{AGB CN Classification}&
  \bf{Reference}\tabularnewline
\hline
   M3&
   8&
   weak&
   \citet{S81}\tabularnewline
\hline
   M4&
   11&
   weak&
   \citet{ISK99}\tabularnewline
\hline
   M5&
   8&
   strong&
   \citet{SN93}\tabularnewline
\hline 
   M13&
   12 &
   weak&
   \citet{S81}\tabularnewline
\hline
   M15&
   2&
   weak&
\citet{L00} \tabularnewline
\hline
   M55&
   10&
   weak&
   \citet{BSH93}\tabularnewline
\hline
   NGC 6752&
   12&
   weak&
   \citet{SN93}\tabularnewline
\hline
   47 Tuc&
   14&
   strong&
   \citet{M78}\tabularnewline
\hline
\end{tabular}
\end{center}
\end{table*}

A vital part in being able to observe significant numbers of AGB stars
is having photometry good enough to separate the AGB from the RGB.
Photometric observations have now reached such high accuracy that
it is becoming feasible to separate the AGB and RGB populations reliably.
In addition to this, wide-field multi-object spectroscopes are now
available. During our literature search we came across some very
high-quality photometric studies. For example, the study of M5 done by \citet{SB04}.
Their set of observations is complete out to 8-10 arc min. They also
tabulate all their stars according to evolutionary status -- and find
105 AGB stars! This represents a sample size increase of \textit{one order of
magnitude}. Further to this we found colour magnitude diagrams
for two more GC candidates that have the required accuracy (and high
AGB star counts). Thus our
current study involves three GCs, one of which appears to have a majority
of CN-strong stars on its AGB (M5) which makes it an important outlier 
that may cause problems for some explanations of the (possible) phenomenon.

Fortuitously, observations of CN bands require only low- to mid-resolution
spectroscopes. This combines well with the fact that a large sample
is required, which is easily achievable with multi-object spectroscopes
which also tend to have lower resolutions. Our proposal also includes 
some (red) HB stars, as this may let us know
which stars reach the HB only, and which stars proceed to the second 
giant branch. RGB stars
will be used as control stars as they are very well studied already
-- and have similar temperatures and luminosities to the AGB stars.
Depending on the quality of the final data we will also attempt to
derive abundances for aluminium and CH (a proxy for carbon).


\section{Discussion}

Assuming for the purpose of discussion that the lack of CN-normal
AGB stars is real, then this is actually the opposite to what we would
expect based on observations at the tip of the RGB. These stars, which are the
precursors to the AGB stars (via the HB), actually tend to become \emph{more} N-dominated
due to `extra mixing' (the results of extra mixing are routinely observed
in Halo field RGB stars as well GC RGB stars -- see eg. \citet{S03}). Thus we would predict an
\emph{increase} in the number of CN-strong AGB stars over the RGB
mean -- rather than a decrease.

\cite{NCF81} proposed two possible explanations to explain the (apparent)
lack of CN-strong stars on the AGB:

\begin{enumerate}

	\item The two populations in NGC 6752 have different He abundances (they
	suggest $\bigtriangleup Y\sim0.05$). This may have come about through
	a merger of two proto-cluster clouds with differing chemical histories
	or through successive generations of stars (ie. self-pollution). The
	He-rich material would also be N-rich. The He-rich stars would then
	evolve to populate the blue end of the HB -- and not ascend the AGB --
	leaving only CN-normal stars to evolve to the AGB. 

	\item Mixing in about half of the RGB stars pollutes their surfaces (increasing
	N) and also increases mass-loss rates, again leading to two separate
	mass groups on the HB. As before, the CN-strong, low mass group does
	not ascend the AGB.

\end{enumerate}

A constraint on the first explanation (for NGC 6752) is that about
\textit{half the mass of the cluster} must be polluted,
as the number of CN-strong and CN-normal stars is roughly equal. As
Norris et al. state, this is not a serious problem for the merger
scenario, as the merging clouds/proto-clusters may very well have
had similar masses. However, due to the constancy of Fe group elements,
the chemical histories of the two clouds/proto-clusters would have
to have been identical with respect to these heavy elements. This
is more difficult to explain since we require a differing chemical histories 
for the light elements.

The self-pollution scenario, whereby a second generation of stars
pollutes the cluster at an early epoch, also needs to satisfy these
two constraints. \citet{FCK04} have explored this
scenario. To maintain the heavy element abundances whilst increasing
N (and other elements) they assume that the cluster does not retain
the ejecta from the second generation supernovae but does retain the
material from the less energetic winds from intermediate mass AGB
stars. Qualitatively AGB stars have a perfect site for the hydrogen
burning needed to produce many of the abundance anomalies in GCs --
the bottom of the convective envelope (so-called `hot-bottom burning').
However, the theoretical study of \citet{FCK04} suggests
that there are actually serious problems for the scenario as the AGB
stars not only produce the N needed but also produce primary carbon
(which is dredged up to the surface). This also alters the sum of
C+N+O significantly which is observed to be (roughly) constant in
GCs. Constraints from other hydrogen burning products also cause this
model to fail.

In light of recent observations on the MS and SGBs of some clusters,
the second explanation by Norris et al. may require some clarification.
As N appears to have a preformation source (as evidenced my MS 
observations), the extra mixing is not
required (although it does still exist). However, the general suggestion
that the differing compositions may affect mass-loss rates and lead
to different mass populations on the HB may be a valid one. 

An important point visible in Table\ref{table1} is that it appears
that there may be variation between the clusters themselves -- some
asymptotic giant branches seem to be CN-strong as opposed to the
majority which appear to be CN-normal. In addition, the fact that most clusters
have a \textit{mix} of CN-strong and CN-normal AGB stars (although usually strongly 
dominated by one population), rather than a homogeneous set, suggests that 
there may be a continuum of CN-strong to CN-normal ratios.
Theories such as those of \cite{NCF81} will have to account
for these points also if the conclusions from observations to date are 
proven correct. Of course, the low sample sizes may be artificially 
complicating the issue.

If there really are substantial abundance differences between the
RGB and AGB then this may also reveal other clues to the GC abundance
anomaly problems (ie. those of the heavier p-capture products - see
\cite{SIK00} for a discussion), and the second parameter problem.

Our study seeks to clarify the understanding of abundance differences
between the various stages of evolution by very significantly increasing 
the amount of information available about the asymptotic branch.  

Finally we note that the AGB stars in question are generally \emph{early} AGB stars
-- they are not thermally pulsing. However, this should have no impact
on the testing for abundance differences as they are not expected
to reduce their surface abundance of nitrogen. Indeed, third dredge-up
on top of preformation pollution and deep mixing would make the issue
even more complex.


\begin{acknowledgements}

The Authors wish to thank the Local and Scientific Organising Committees
for making the Eighth Torino Workshop on Nucleosynthesis in AGB Stars
a rewarding event. In particular Simon Campbell would like to thank the organisers for
the complete support given to the PhD students attending the workshop. 
He would also like to thank the Australian Astronomical Society for the
travel funding received.

\end{acknowledgements}


\bibliographystyle{aa}

\begin{thebibliography}{}

\bibitem[Briley et al.(1993)]{BSH93}
Briley, M. M.; Smith, G. H.; Hesser, J. E.; Bell, R. A., 1993, AJ, 106, 142
	
\bibitem[Cannon et al.(1998)]{CCB98}
Cannon, R. D., Croke, B. F. W., Bell, R. A., Hesser, J. E., Stathakis, R. A., 1998, MNRAS, 298, 601

\bibitem[Fenner et al.(2004)]{FCK04}
Fenner, Y., Campbell, S., Karakas, A. I., Lattanzio, J. C., Gibson, B. K., 2004, MNRAS, 353, 789

\bibitem[Gratton et al.(2000)]{GSC00}
Gratton, R. G., Sneden, C., Carretta, E., Bragaglia, A., 2000, A\&A, 354, 169

\bibitem[Ivans et al.(1999)]{ISK99}
Ivans, I.I., Sneden, C., Kraft, R.P., Suntzeff, N.B., Smith, V.V., Langer, G.E., Fulbright, J.P., 1999, AJ, 118, 1273I

\bibitem[Kraft et al.(1992)]{KSL92}
Kraft, R. P., Sneden, C., Langer, G. E., Prosser, C. F., 1992, ApJ, 104, 645

\bibitem[Lee(2000)]{L00}
Lee, S. G., 2000, JKAS, 33, 137 

\bibitem[Mallia(1978)]{M78}
Mallia, E. A., 1978, A\&A, 70, 115

\bibitem[Norris et al.(1981)]{NCF81}
Norris, J., Cottrell, P. L., Freeman, K. C., 1981, ApJ, 244, 205

\bibitem[Pilachowski et al.(1996a)]{PSK96a}
Pilachowski, C. A., Sneden, C., Kraft, R. P., 1996a, AJ, 111, 1689

\bibitem[Pilachowski et al.(1996b)]{PSK96b}
Pilachowski, C. A., Sneden, C., Kraft, R. P., Langer, G. E., 1996a, AJ, 112, 545

\bibitem[Sandquist \& Bolte(2004)]{SB04}
Sandquist, E. L., Bolte, M, 2004, ApJ, 611, 323

\bibitem[Shetrone (2003)]{S03}
Shetrone, M. D., 2003, ApJ, 585L, 45

\bibitem[Smith \& Norris(1993)]{SN93}
Smith, G. H., Norris, J, E., 1993, AJ, 105, 173

\bibitem[Sneden et al.(2000)]{SIK00}
Sneden, C., Ivans, I. I., Kraft, R. P., 2000, MmSAI, 71, 657

\bibitem[Suntzeff(1981)]{S81}
Suntzeff, N. B., 1981, ApJS, 47, 1

\bibitem[Suntzeff \& Smith(1991)]{SS91}
Suntzeff, N. B., Smith, V. V., 1991, ApJ, 381, 160

\end{thebibliography}

\end{document}